\documentclass[twocolumn]{aastex61}

\received{2017 September 13}
\revised{2017 December 20}
\accepted{2017 December 24}

\usepackage{graphicx}
\usepackage{amssymb}
\usepackage{amsmath}
\usepackage{epstopdf}
\shorttitle{Molecular reconnaissance of the $\beta$ Pictoris gas disk}
\shortauthors{Matr\`a et al.}

\begin{document}


\title{Molecular reconnaissance of the $\beta$ Pictoris gas disk with the SMA: \\ a low HCN/(CO+CO$_2$) outgassing ratio and predictions for future surveys}

\author[0000-0003-4705-3188]{L. Matr\`a}
\altaffiliation{Submillimeter Array (SMA) Fellow}
\affil{Harvard-Smithsonian Center for Astrophysics, 60 Garden Street, Cambridge, MA 02138, USA}
\email{luca.matra@cfa.harvard.edu}
\author{D. J. Wilner}
\affil{Harvard-Smithsonian Center for Astrophysics, 60 Garden Street, Cambridge, MA 02138, USA}
\author{K. I. \"{O}berg}
\affil{Harvard-Smithsonian Center for Astrophysics, 60 Garden Street, Cambridge, MA 02138, USA}
\author{S. M. Andrews}
\affil{Harvard-Smithsonian Center for Astrophysics, 60 Garden Street, Cambridge, MA 02138, USA} 
\author{R. A. Loomis}
\affil{Harvard-Smithsonian Center for Astrophysics, 60 Garden Street, Cambridge, MA 02138, USA}
\author{M. C. Wyatt}
\affil{Institute of Astronomy, University of Cambridge, Madingley Road, Cambridge CB3 0HA, UK}
\author{W. R. F. Dent}
\affil{ALMA Santiago Central Offices, Alonso de Cordova 3107, Vitacura, Santiago, Chile}

\begin{abstract}
The exocometary origin of CO gas has been confirmed in several extrasolar Kuiper belts, with CO ice abundances consistent with Solar System comets. We here present a molecular survey of the $\beta$ Pictoris belt with the Submillimeter Array (SMA), reporting upper limits for CN, HCN, HCO$^+$, N$_2$H$^+$ and H$_2$CO, as well as for H$_2$S, CH$_3$OH, SiO and DCN from archival ALMA data. Non-detections can be attributed to rapid molecular photodissociation due to the A-star's strong UV flux. CN is the longest-lasting and most easily detectable molecule after CO in this environment. We update our NLTE excitation model to include UV fluorescence, finding it plays a key role in CO and CN excitation, and use it to turn the SMA CN/CO flux ratio constraint into an upper limit of $<2.5$\% on the HCN/(CO+CO$_2$) ratio of outgassing rates.
This value is consistent with, but at the low end of, the broad range observed in Solar System comets. 
If sublimation dominates outgassing, then this low value may be caused by decreased outgassing for the less volatile molecule HCN compared to CO. If instead UV photodesorption or collisional vaporization of unbound grains
dominates outgassing, then this low ratio of rates would imply a low ice abundance ratio, which would in turn indicate a variation in cometary cyanide abundances across planetary systems.
To conclude, we make predictions for future molecular surveys and show that CN and HCN should be readily detectable with ALMA around $\beta$ Pictoris for Solar-System-like exocometary compositions.


\end{abstract}






\section{Introduction}
\label{sect:intro}

Sensitive searches with the Atacama Large Millimeter/submillimeter Array (ALMA) are steadily adding to the number of disks with dust levels typical of debris disks but hosting detected CO gas, now comprising 15 systems \citep{Zuckerman1995,Moor2011,Dent2014,Moor2015,Marino2016, Lieman-Sifry2016,Greaves2016,Matra2017b,Marino2017a,Moor2017}. All but two of these are relatively young ($\sim$10-100 Myr-old), which would naively suggest that the presence of CO is related to the protoplanetary era of disk evolution. However, the recent detections of CO around the 440 Myr-old star Fomalhaut \citep{Matra2017b} and the 1-2 Gyr-old star $\eta$ Corvi \citep{Marino2017a}, as well as the tentative detection around the 125-130 Myr-old eclipsing binary HD23642 \citep{Pericaud2017}, show that at least in some of the systems the CO cannot be primordial. Instead, the wide range of system ages advocates for CO replenishment through gas release from exocometary ices.

This exocomet scenario has now been confirmed as the origin of the observed CO gas in four systems, including relatively young stars $\beta$ Pictoris \citep{Dent2014,Matra2017a} and HD181327 \citep{Marino2016}, as well as Fomalhaut and $\eta$ Corvi. While the exact CO release mechanism itself remains unknown, the similarly clumpy CO and dust distributions in the disk around $\beta$ Pictoris suggest that collisions play an important role \citep{Dent2014}. A collisional cascade is expected to result in efficient gas release as long as the smallest grains blown out of the system by radiation pressure from the central star cannot retain their ice \citep{Matra2017b}. In this case, assuming steady state in both the collisional cascade and in the production/photodestruction of gas allows us to extract the CO ice fraction in exocomets from simple observables \citep{Zuckerman2012,Matra2015,Matra2017b}. This is a particularly powerful tool to probe the bulk volatile composition of exocomets, allowing us to place Solar System comets in the larger context of planetary systems around stars with different properties. 


In the debris disk systems where an exocometary origin of CO has been confirmed (two A- and one F-type stars), measured CO mass fractions of exocomets are consistent with Solar System comets \citep{Marino2016,Matra2017a,Matra2017b}. This consistency suggests similar comet formation conditions in the original protoplanetary disks and the Solar Nebula. The next step in investigating this similarity requires to both expand this sample of exocometary CO abundance measurements and probe relative ice abundances through detection of other molecules. As we discuss in this paper (\S\ref{sect:disc}), detection of other molecules can uniquely shed light on both the exocometary outgassing mechanism (where outgassing refers to any gas release process from solid bodies of any size) and the chemical history of cometary volatiles.

Here we present the first molecular survey of the exocometary gas disk around $\beta$ Pictoris. In \S\ref{sect:obs}, we describe new SMA and archival ALMA observations of CO, CN, HCN, HCO$^+$, N$_2$H$^+$, H$_2$CO, H$_2$S, CH$_3$OH, SiO and DCN. The continuum and CO J=2-1 detections with the SMA as well as SMA and ALMA upper limits for all other molecules are presented in \S\ref{sect:res}. We interpret the observations with a steady state exocometary release model in \S\ref{sect:anal}. In particular, we focus on CN and CO as the longest-lived observable gas-phase species in the disk, and describe the connection between the constraint on the observed CN/CO flux ratio upper limit and an upper limit in the HCN/(CO+CO$_2$) ratio of outgassing rates. We compare this ratio with measurements from Solar System comets and discuss the implications of this comparison on the release mechanism and/or on the HCN/(CO+CO$_2$) ice abundance in the exocomets in \S\ref{sect:disc}. Finally, we use our exocometary release model to make predictions for molecular detection in future surveys from far-IR to radio wavelengths, and conclude with a summary in \S\ref{sect:concl}.

\section{Observations}
\label{sect:obs}

\begin{deluxetable*}{lcc}
\tablewidth{0pt}
\tablecaption{SMA Observational Parameters}
\tablehead{
\colhead{Parameter} & \colhead{2014 Nov 11} & \colhead{2015 Dec 15}}
\startdata
No. Antennas & 7 & 8 \\
$\tau_{\rm 225~GHz}$ & 0.07 & 0.04 \\
Pointing center &
\multicolumn{2}{c}{
  R.A. $05^{h}47^{m}17\fs0877$, Dec. $-51^{h}03^{m}59\fs44$
  (J2000) } \\
Min/Max baseline  & 6 to 67 meters & 6 to 67 meters \\
Gain Calibrators  & J0522-364 (2.7 Jy) & J0522-364 (4.9 Jy) \\
                  & J0538-440 (1.7 Jy) & J0538-440 (1.5 Jy) \\
Passband Calibrator & 3C84, Uranus & 3C84, Uranus \\
Flux Calibrator    & Callisto & Uranus \\
Spectral Lines & ~ & ~ \\
~~transition, frequency  & HCN J=3--2, 265.88643 GHz
                             & CO J=2-1, 230.538 GHz  \\
~~                       & HCO$^+$ J=3--2, 267.55763 GHz
                             & CN N= 2-1, 226.87478 GHz \\
~~                       & N$_2$H$^+$ J=3--2, 279.51176 GHz
                             & (J=5/2-3/2 component) \\
~~                       & H$_2$CO J=$4_{14}-3_{13}$, 281.52693 GHz
                             &  \\
Primary Beam FWHM & $43''$ & $55''$ \\
Synthesized Beam FWHM\tablenotemark{a}
  & $7\farcs7 \times 3\farcs3$, p.a. $12^{\circ}$
  & $8\farcs7 \times 3\farcs7$, p.a. $-10^{\circ}$ \\
rms noise (line images)          & 0.09 Jy beam$^{-1}$ @ 265.9 GHz & 0.06 Jy beam$^{-1}$ @ 230.5 GHz \\
          & (for 0.92 km/s channels) &  (for 1.06 km/s channels)\\
rms noise (continuum image)      &  1.4 mJy beam$^{-1}$ & 0.57 mJy beam$^{-1}$ \\
\enddata
\tablenotetext{a}{for natural weighting plus a
$3''$ Gaussian taper in the east-west direction}
\label{tab:obs_sma}
\end{deluxetable*}

\subsection{SMA}
\label{sect:obssma}
We observed the $\beta$~Pictoris debris disk with the Submillimeter Array
(SMA) on Mauna Kea, Hawaii in two tracks in a compact antenna configuration
to search for line emission from a suite of molecular species.
Table~\ref{tab:obs_sma} provides a summary of the observational parameters.
The first track was obtained on 2014 November 11 using 7 antennas.
The ASIC correlator was configured with 128 channels (0.8125 MHz spacing)
in each of 48 chunks that together spanned $\pm(4-8)$~GHz from an LO frequency
of 273.732 GHz. This setup was aimed at simultaneous coverage of spectral
lines of HCN and HCO$^+$ (LSB) and N$_2$H$^+$ and H$_2$CO (USB).
The second track was obtained on 2015 December 15 with 8 antennas.
The ASIC correlator was again configured with 128 channels (0.8125 MHz
spacing) in each of 48 chunks, spanning $\pm(4-8)$~GHz from an LO frequency
of 221.654 GHz. In addition, the partly deployed SWARM correlator
(operating at $8/11$ speed) provided an additional 2 chunks of 16384 channels
each ($0.8125/8$~MHz spacing) that spanned $\pm(8-12)$~GHz from the LO
frequency with a gap centered at $10$~GHz of width $1.15$~GHz.  This setup
was aimed at simultaneous coverage of spectral lines of CO and CN (USB).

The southern location of $\beta$~Pictoris (Dec. $-51\degr$) makes it a
challenging target for the SMA, since it is available only at high airmass,
rising above an elevation of 15 degrees for only about 3 hours.
However, for both of these tracks the weather conditions were very good,
with stable atmospheric phase and low opacity at 225~GHz as measured by
the nearby tipping radiometer. The basic observing sequence consisted of a
loop of 2 minutes each on the quasars J0522-364 and J0538-440 and 8 minutes
on $\beta$~Pictoris.  Passband calibration was obtained with observations
of a combination of available strong sources. The absolute flux scale for
the two tracks were set using observations of Callisto and Uranus, with
an estimated accuracy of 20\%.  All of the calibration was performed
using the {\tt MIR\footnote{\url{https://www.cfa.harvard.edu/~cqi/mircook.html}}} software package. Continuum and spectral line imaging
and clean deconvolution were done in the {\tt miriad} package.

The 2014 dataset presented a systematic astrometric offset of $\sim 2.4\arcsec$ (a fraction of the $8\farcs7\times3\farcs7$ beam size) with respect to location of the star. We corrected this offset using the 274 GHz continuum image from the same 2014 dataset, by measuring the centroid of continuum emission and spatially shifting the image to ensure that this centroid lies at the expected stellar location.     

\subsection{ALMA}
\label{sect:obsalma}
We extracted ALMA Band 6 observations of $\beta$ Pictoris from the archive (project number 2012.1.00142.S) in order to search for line emission from several additional molecular species. In summary, observations were performed between 2013 and 2015 with the 12-m array in a compact and extended configuration, as well as with the Atacama Compact Array (ACA), with total on-source times of 28, 114, and 50 mins, respectively. We direct the reader to \citet{Matra2017a} for a detailed description of the calibration and CO imaging procedures. Two of the spectral windows were set in frequency division mode and allowed us to test for the presence of spectral line emission from H$_2$S (2$_{2,0}$-2$_{1,1}$, 216.710 GHz), CH$_3$OH (5$_{1,4}$-4$_{2,2}$, 216.946 GHz) SiO (J=5-4, 217.105 GHz) and DCN (J=3-2, 217.239 GHz), as well as CO (J=2-1, 230.538 GHz). We here repeated the imaging step for all molecules in the same way as previously done for the CO transition \citep{Matra2017a}; this resulted in 4 line datasets with a native channel spacing of 488.281 kHz (corresponding to a resolution of 1.35 km/s at the rest frequency of the SiO line). To allow direct comparison with CO (\S\ref{sect:filtering}), we applied a taper to the visibilities in order to produce datacubes with exactly the same synthesized beam size ($0\farcs30\times0\farcs26$, or 13.4$\times$10.7 au at the distance of $\beta$ Pictoris) and position angle (p.a., -83$\fdg$9) as the CO dataset.

\section{Results}
\label{sect:res}

\subsection{SMA continuum and CO J=2-1 detections}

\begin{figure*}
 \hspace{15mm}
  \includegraphics*[scale=0.49]{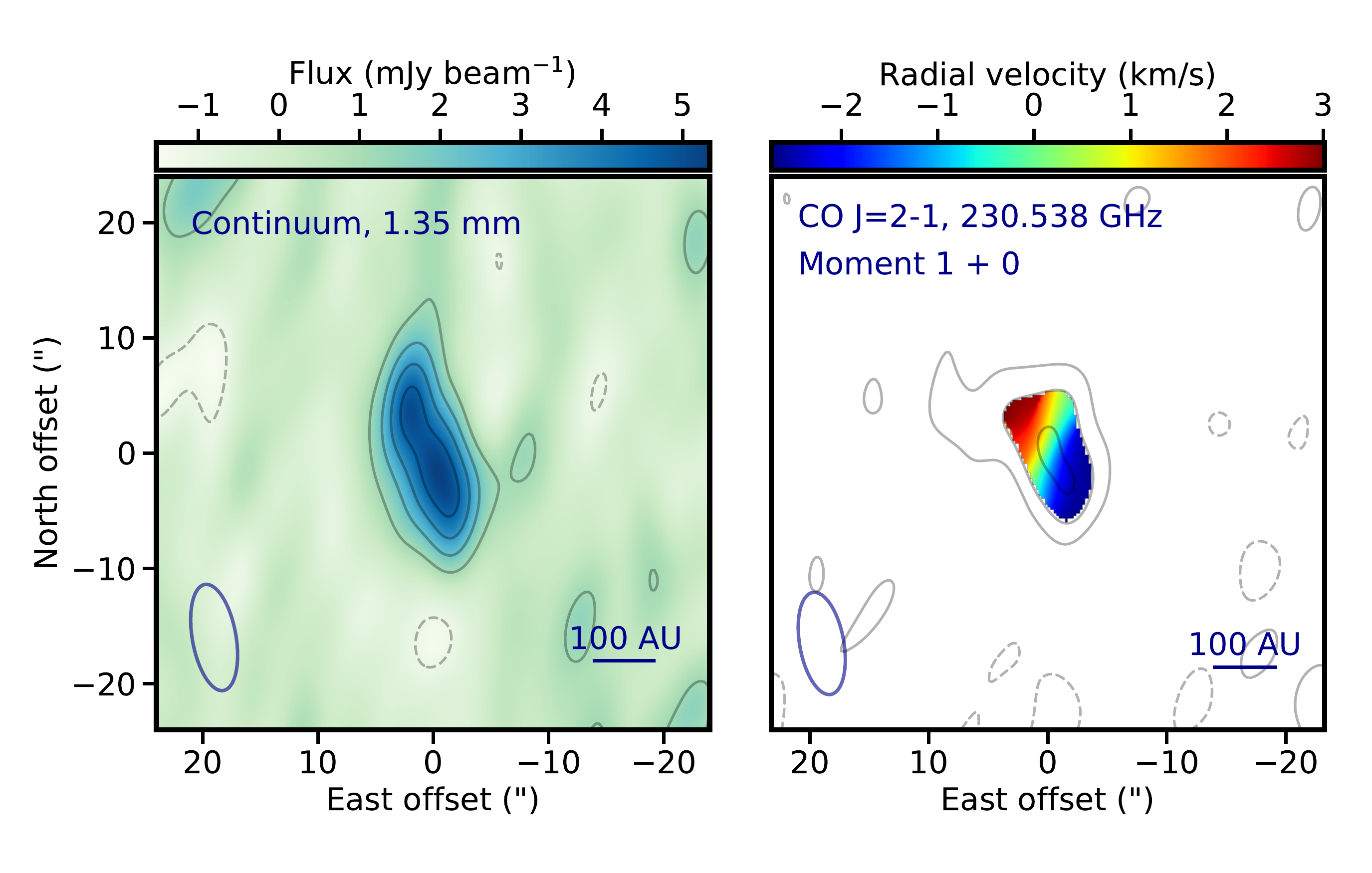}
\vspace{-6mm}
\caption{\textit{Left}: SMA 1.35 mm continuum image of the $\beta$ Pictoris disk. Contours represent [-2,2,4,6,8] times the rms noise level of 0.57 mJy beam$^{-1}$. \textit{Right}: Contours show the moment 0 (spectrally integrated) map of the CO J=2-1 line, at [-2,2,4,6] times the rms noise level of 0.18 Jy km s$^{-1}$ beam$^{-1}$. The colour scale (moment 1 map) represents the velocity field of the disk, i.e. the centroid velocity of the line (with respect to the star) at each spatial location. In both panels, the ellipse in the bottom left represents the synthesized beam.} 

\label{fig:biplot}
\end{figure*}

Continuum emission from dust within the $\beta$ Pictoris disk is detected in both SMA datasets at 222 and 274 GHz (1.35 and 1.09 mm). Figure~\ref{fig:biplot} (left) shows the SMA image with the highest signal-to-noise ratio (1.35 mm), presenting a double-peaked morphology consistent with the 1.09 mm dataset, as well as previous SMA 236 GHz \citep[1.27 mm,][]{Wilner2011} and ALMA 339 GHz \citep[0.89 mm,][]{Dent2014} datasets, interpreted as a belt viewed almost exactly edge-on.
Flux density measurements from the SMA observations at 222 and 274 GHz are $21.0\pm3.0$~mJy and $25.9\pm6.4$~mJy, respectively
(obtained by fitting a Gaussian to the visibilities at baselines shorter than 30~k$\lambda$). These values are consistent with 
the ALMA measurement at 339 GHz ($60\pm6$~mJy), given the large uncertainties and steep spectral index of the emission \citep[$\alpha\sim$2.81,][]{Ricci2015}. The previously reported SMA 236 GHz continuum measurement remains anomalously low ($13\pm1.4$~mJy), perhaps resulting from calibration challenges at very low elevations.     

CO J=2-1 emission is clearly detected in the SMA channel maps, centered around a velocity of $\sim 1.4$ km/s with respect to the rest frequency of 230.538 GHz in the Local Standard of Rest (LSR) reference frame. We created a moment 0 image (contours in Figure~\ref{fig:biplot}, right) by spectrally integrating the datacube within $\pm 7$ km/s of the stellar velocity (the latter being consistent with the centroid velocity reported above). We used this image to create a spatial mask large enough to include all of the emission from the disk. Then, we applied this mask to the original datacube as a spatial filter; in other words, we spatially integrated emission within this mask. This produced the 1D spectrum shown in Figure~\ref{fig:allmolspec} (top left, lilac line). 

The CO line profile is spectrally resolved and appears asymmetric; the enhancement at blueshifted radial velocities is consistent with previous ALMA observations showing that the majority of CO emission originates in a clump on the south-west (SW) side of the edge-on disk, where this SW side is rotating towards Earth \citep{Dent2014,Matra2017a}. This is confirmed by the moment 1 map (color map in Figure~\ref{fig:biplot} right), which shows the line-of-sight velocity centroid of CO emission at spatial locations where the disk is detected (at the 4$\sigma$ level).  This centroid was obtained by fitting Gaussians to spectra at all spatial locations. Within the measurement uncertainties, the CO J=2-1 line shape and integrated line flux are consistent with the previous ALMA measurements (see ALMA CO spectrum as the lilac line in Figure~\ref{fig:allmolspec}, top right, as well as CO line fluxes in Table \ref{tab:mols}).

\subsection{Upper limits from molecular survey: spectro-spatial filtering using CO as a template}
\label{sect:filtering}

\begin{figure*}
 \centering
  \includegraphics*[scale=0.5]{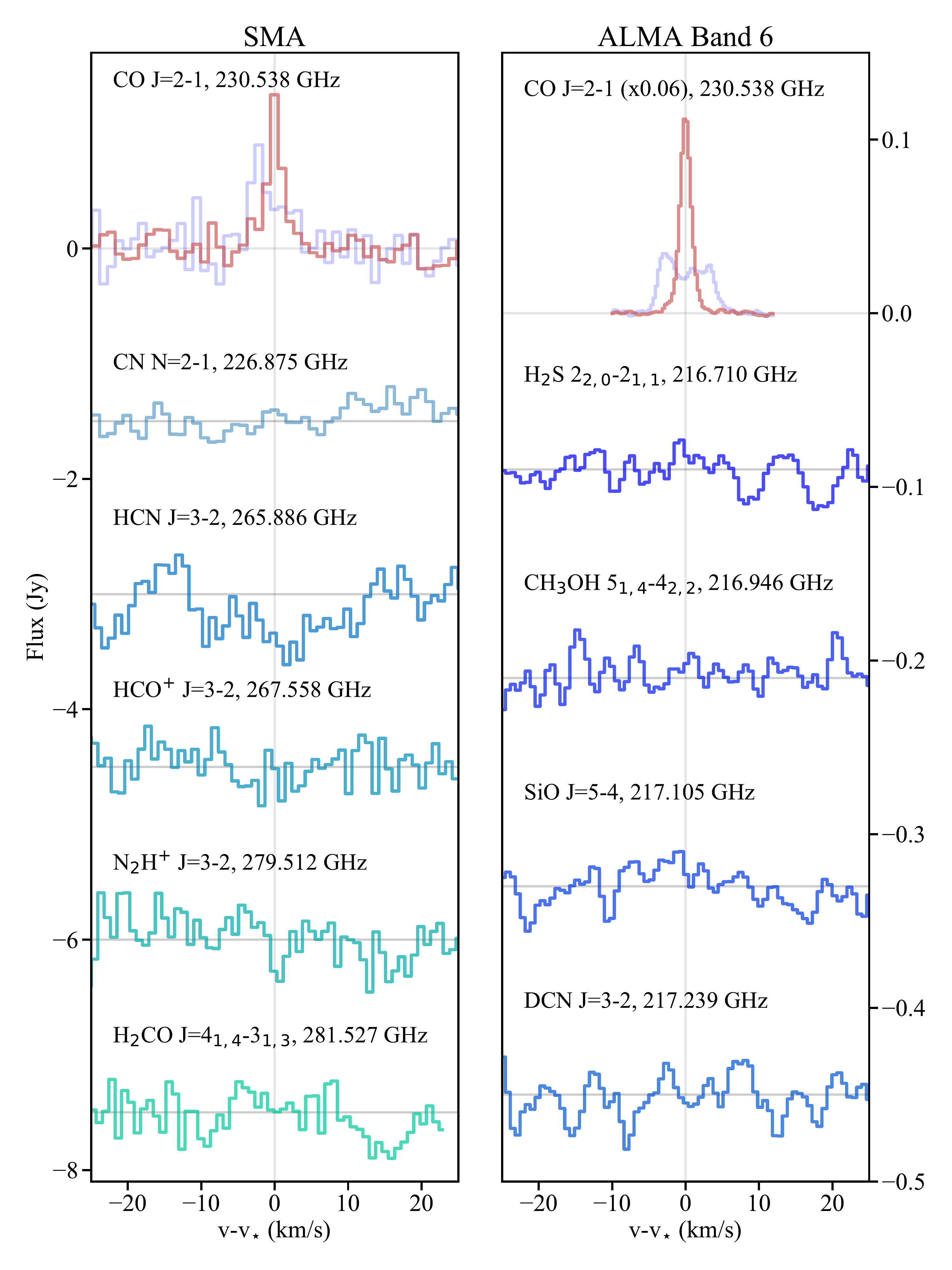}
\vspace{-4mm}
\caption{Spectra of the lines targeted in our SMA survey (left) and in archival ALMA data (right). For the CO J=2-1 line (top row), we show the effect of our spectro-spatial filtering technique (see \S\ref{sect:filtering}). Lilac represents the spectrum of the disk spatially integrated over a mask covering all significant emission from the disk. The red line is the result of our technique, which additionally shifts spectra to the zero velocity at each spatial location, leading to a significant S/N boost at the zero velocity. For all other species, we only display spectra after filtering in the same way as done for the CO. Note the different flux scale of the two columns, and of the CO J=2-1 line observed by ALMA with respect to all other lines. Line fluxes and upper limits are reported in Table \ref{tab:mols}.}
\label{fig:allmolspec}
\end{figure*}

No significant signal is detected in the SMA datacubes created around the spectral location corresponding to the CN N=2-1, J=$\frac{5}{2}$-$\frac{3}{2}$ line (composed of the 3 blended F=$\frac{7}{2}$-$\frac{5}{2}$, $\frac{5}{2}$-$\frac{3}{2}$ and $\frac{3}{2}$-$\frac{1}{2}$ hyperfine transitions), HCN J=3-2, HCO$^{+}$ J=3-2 , N$_2$H$^{+}$ J=3-2 and H$_2$CO J=4$_{1,4}$-$3_{1,3}$ lines. Similarly, no significant emission is observed in the ALMA datacubes around the H$_2$S 2$_{2,0}$-2$_{1,1}$, CH$_3$OH 5$_{1,4}$-4$_{2,2}$, SiO J=5-4 and DCN J=3-2  lines.

To optimize detectability, we implement a spectro-spatial filtering technique similar to that introduced in \citet{Matra2015} and subsequently applied to achieve CO detections in the HD181327, $\eta$~Corvi and Fomalhaut debris disks \citep{Marino2016, Marino2017a, Matra2017b}. The method involves choosing a prior on the spectro-spatial distribution of the gas emission, and using this prior to select pixels/channel combinations where emission is expected. This maximizes the signal-to-noise ratio by effectively removing noise-dominated pixels/channels. 

We demonstrate this technique by applying it first to the CO J=2-1 line already detected without need for filtering by both the SMA and ALMA. In practice, the process involves two steps. First, we apply spectral filtering by shifting all 1D spectra at each spatial location in the datacube by the negative of the centroid CO velocity displayed in the moment 1 map (Figure~\ref{fig:biplot}, right). This causes the spectra to align such that the bulk of the CO emission now lies at the stellar velocity at all spatial locations. Second, we apply spatial filtering by spatially integrating over the previously created mask covering the region where CO emission is detected. This produces the CO spectra shown in 
Figure~\ref{fig:allmolspec} (top, red line) for the SMA (left) and ALMA (right) observations. 
This technique significantly improves the signal-to-noise of the detection with respect to the CO spectra directly extracted from the spectral image cube (lilac lines). The improvement is clearer for the ALMA dataset since the line is better resolved both spectrally and spatially, increasing the number of spatial locations that are spectrally independent of one another. In other words, if the emission is better resolved spatially, then the spectral lines at each spatial location will be narrower, and the spectral shifting will produce a better alignment and consequential signal--noise ratio boost. 

For simplicity, we here assume emission from all molecules to have similar spectro-spatial distributions of emission as observed for the CO molecule (though see caveats in \S\ref{sect:pred}). This allows us to use the observed CO emission as a template for the spectro-spatial filtering technique. Figure~\ref{fig:allmolspec} shows the resulting filtered spectra for all observed molecules. There are no significant detections of any molecular lines other than CO. We measure the filtered integrated flux upper limits by multiplying the filtered spectrum's root mean square (RMS, measured away from the line frequency) by the number of independent spectral channels covering the line width across which CO is detected. The $3\sigma$ limits derived are listed in Table \ref{tab:mols}. As this assumes that all molecules have the same spectro-spatial distribution as CO, we also calculate upper limits in a more agnostic way by spatially integrating the unfiltered data cubes over regions where the mm continuum is detected, and spectrally integrating between $\pm$5 km/s of the stellar velocity (values in parentheses in Table 2). Whether these are more or less stringent than the CO-filtered limits depends on both the on-sky area over which the continuum versus CO are detected, and on the spectral width of the CO line before versus after spectral filtering, which differs for the SMA and ALMA datasets (see top row in Fig. \ref{fig:allmolspec}).


\begin{deluxetable}{cccccc}
\tabletypesize{\scriptsize}
\tablecaption{Observations of molecular line emission around $\beta$ Pictoris at millimeter wavelengths \label{tab:mols}}
\tablewidth{0pt}
\tablehead{
\colhead{Species} & \colhead{Transition} & \colhead{Freq.} & \colhead{Line Flux} & \colhead{Instr.} &
\colhead{Ref.} \\
\colhead{ } & \colhead{ } & \colhead{(GHz)} & \colhead{(Jy km/s)} & \colhead{ } &
\colhead{ }
}
\startdata
H$_2$S & 2$_{2,0}$-2$_{1,1}$ & 216.710 & $\leqslant$0.10 (0.18) & ALMA & 1 \\
CH$_3$OH & 5$_{1,4}$-4$_{2,2}$ & 216.946 & $\leqslant$0.11 (0.13) & ALMA & 1 \\
SiO & J=5-4 & 217.105 & $\leqslant$0.11 (0.16) & ALMA & 1 \\
DCN & J=3-2 & 217.239 & $\leqslant$0.13 (0.16) & ALMA & 1 \\
CO & J=2-1 & 230.538 & 4.5$\pm$0.5 & ALMA & 2 \\
 & & & 3.9$\pm$0.8 & SMA & 1 \\
  & J=3-2 & 345.796 & 5.8$\pm$0.6 & ALMA & 2, 3 \\
CN & N=2-1 & 226.875 & $\leqslant$1.0 (1.2) & SMA & 1 \\
HCN & J=3-2 & 265.886 & $\leqslant$2.0 (1.4) & SMA & 1 \\
HCO$^{+}$ & J=3-2 & 267.558 & $\leqslant$1.5 (1.1) & SMA & 1 \\
N$_2$H$^{+}$ & J=3-2 & 279.512 & $\leqslant$1.7 (1.2)  & SMA & 1 \\
H$_2$CO & J=4$_{1,4}$-$3_{1,3}$ & 281.527 & $\leqslant$1.6 (1.2) & SMA & 1 \\
\enddata


\tablecomments{For CN, the upper limits refer to the J=$\frac{5}{2}$-$\frac{3}{2}$ transition. Line flux upper limits reported are at the 3$\sigma$ level, after spectro-spatial filtering. Values in parentheses are upper limits obtained from spatial integration over the region where the continuum is detected at a level $>2\sigma$ in each respective dataset.} References: (1) This work; (2) \citet{Matra2017a}; (3) \citet{Dent2014}

\end{deluxetable}

\section{Analysis}
\label{sect:anal}
We investigate how to connect the line flux upper limits measured in \S\ref{sect:res} to exocometary ice compositions relative to CO, the only molecular species detected so far in exocometary gas disks. In particular, we aim to understand the significance of these upper limits in relation to Solar System cometary compositions. We put a special emphasis on the CN/CO ratio upper limit and its consequences for HCN and CO in $\beta$ Pictoris' exocometary ices, for reasons which will become apparent in \S\ref{sect:lifetimes}.

\subsection{Expanding the steady state cometary release model}
\label{sect:expandmodel}
In order to understand the factors that affect remote detection of molecular species released in the gas phase from exocometary ices, we
use the steady state exocometary gas release model of \citet{Matra2015,Matra2017b} and  expand it to include molecular species other than CO. Within the framework of this model, the mass production rate $\dot{M_{\rm i}^+}$ of molecular species i will equal its destruction rate $\dot{M_{\rm i}^-}$,

\begin{equation}
\dot{M_{\rm i}^+}=m_{\rm i}\dot{N_{\rm i}^+}=\dot{M_{\rm i}^-}=\frac{M_{\rm i}}{\tau_{\rm i}^-},
\end{equation}
where $m_{\rm i}$ is its molecular mass in kg, $\dot{N_{\rm i}^+}$ its production rate in number of molecules per second,  $M_{\rm i}$ its observed gas mass, and $\tau_{\rm i}^-$ its destruction timescale in the gas phase, which we here assume to be dominated by photodissociation, neglecting any other gas phase chemical reactions ($\tau_{\rm i}^-=\tau_{\rm i, phd}^-$). 

Then, the gas mass of species i we would expect to observe relative to CO is 
\begin{equation}
\frac{M_{\rm i}}{M_{\rm CO}}=\frac{\tau_{\rm i, phd}^-m_{\rm i}}{\tau_{\rm CO, phd}^-m_{\rm CO}}\frac{\dot{N_{\rm i}^+}}{\dot{N_{\rm CO}^+}}.
\end{equation}
If CO and species i are released in the gas at the same rate \textit{per molecule} (an assumption that we will revisit in \S\ref{sect:depl}) we can directly link an observed gas mass ratio to an ice abundance ratio in the exocomets. Furthermore, this relation shows the crucial impact of the photodissociation timescale on the observable mass of exocometary gas species (see \S\ref{sect:lifetimes}).

The next step is to link the molecular gas mass to line fluxes for given transitions as observed in this work. Following e.g. Eq. 2 in \citet{Matra2015}, and assuming optically thin transitions (see \S\ref{sect:pred}),
\begin{equation}
\label{eq:releaseratetoflux}
\frac{F_{\nu_{\rm u\rightarrow l, i}}}{F_{\nu_{\rm u\rightarrow l, CO}}}=\frac{ A_{\rm u\rightarrow l, i}}{A_{\rm u\rightarrow l, CO}}\frac{x_{\rm u, i}}{x_{\rm u, CO}}\frac{\tau_{\rm i, phd}^-}{\tau_{\rm CO, phd}^-}\frac{\dot{N_{\rm i}^+}}{\dot{N_{\rm CO}^+}}, 
\end{equation}
where, for both CO and species i, $F_{\nu_{\rm u\rightarrow l}}$ are integrated line fluxes in Jy km/s, $A$ and $x_{\rm u}$ are Einstein A coefficients and upper level fractional populations of the observed transition. 
We note that the steady state relation in Eq. \ref{eq:releaseratetoflux} remains the same for daughter molecular species such as CN and OH, as long as their production is dominated by a single parent species (such as HCN and H$_2$O). 

\subsection{The effect of photodissociation: lifetime of Solar System cometary molecules}
\label{sect:lifetimes}

\begin{figure*}
\vspace{-0mm}
 \hspace{-2mm}
  \includegraphics*[scale=0.65]{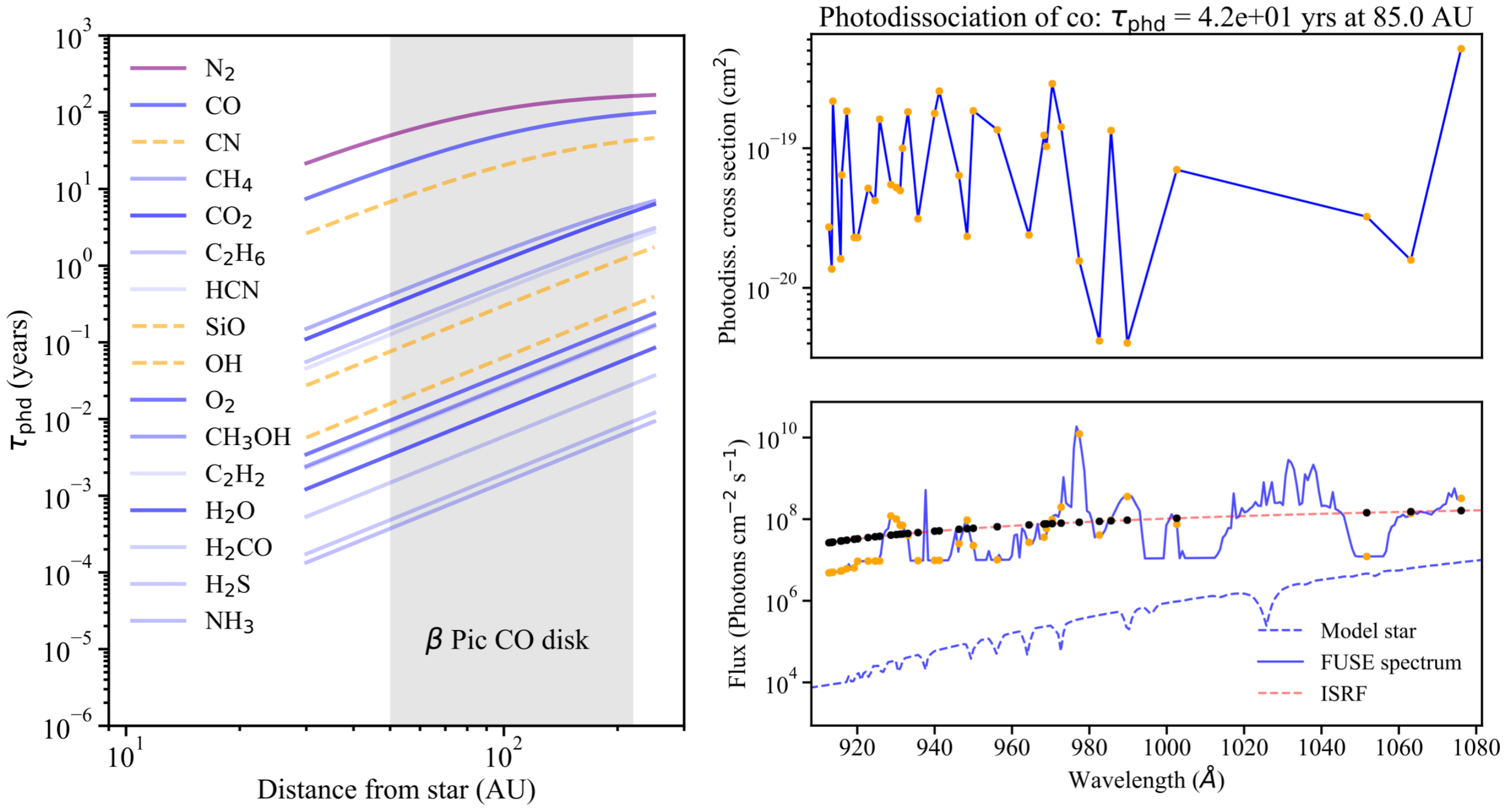}
\vspace{-5mm}
\caption{\textit{Left:} Unshielded photodissociation timescales of molecular gas species observed in Solar System comets. Species are listed in order of their timescale, from longest lived to shortest lived, and are therefore in the same vertical order as the corresponding lines (though note that the C$_2$H$_2$ line overlaps with the CH$_3$OH line). The transparency of solid lines is set to indicate their log-scale abundance in Solar System cometary ice with respect to H$_2$O, darker for more abundant species. The purple line is for N$_2$, by far the least abundant of all listed species from measurements in comet 67P \citep{Rubin2015}. Orange dashed lines represent daughter species (CN, OH) or species that we observed but are not detected in Solar System comets (SiO). \textit{Right:} Example of our derivation of photodissociation timescales for CO; the top panel shows cross sections at relevant wavelengths obtained from the \citet{Heays2017} database. The bottom panel shows the mean intensity felt by a molecule at 85 AU in the $\beta$ Pic disk (assuming the gas to be optically thin to UV photons). We show contributions from both the ISRF and the star, highlighting the difference between the observed stellar flux of $\beta$ Pictoris and that expected from a model of the same spectral type as the star.
}
\label{fig:timescales}
\end{figure*}

In Figure~\ref{fig:timescales} (left), we present unshielded photodissociation timescales for molecules observed around $\beta$ Pic as well as other species observed to be abundant in Solar System comets. We do not include HCO$^+$ and N$_2$H$^+$ as these ions, if present at all, would likely recombine through free electrons faster than they photodissociate; taking the example of HCO$^+$, using an electron density of 10$^2$-10$^3$ cm$^{-3}$ \citep{Matra2017a}, temperatures between 20 and 100 K and reaction rates from the KIDA \footnote{\url{http://kida.obs.u-bordeaux1.fr}} database \citep{Wakelam2012}, its recombination timescale is 0.2 to 5 days, much shorter than the $3.7\times10^3$ yr photodissociation timescale at 85 AU from the star.

Photodissociation timescales for all molecules were calculated using photodissociation cross sections from \citet[][]{Heays2017}\footnote{\url{http://home.strw.leidenuniv.nl/~ewine/photo/}} and references therein, neglecting isotope-selective photodissociation for rarer isotopologues. The radiation field includes both the observed spectrum of the $\beta$ Pic star and the interstellar radiation field (ISRF). The stellar spectrum (A. Brandeker, priv. comm.) was obtained by superimposing observed spectra from HST/STIS \citep{Roberge2000} and FUSE \citep{Bouret2002,Roberge2006} on the photospheric model spectrum of \citet{Fernandez2006}. For the ISRF we adopted the formulation of \citet{Draine1978} with the long wavelength expansion of \citet{vanDishoeck2006}.

We note that the $\beta$ Pictoris star is surprisingly active for an A star, with X-ray observations presenting evidence for thermal emission from a cool corona \citep{Hempel2005, Gunther2012}, as well as UV chromospheric emission \citep{Deleuil2001, Bouret2002}. The latter effectively causes a considerable UV flux `excess' compared to a typical main sequence A-star; this acts to shorten the survival lifetime of molecules against photodissociation, particularly for molecules whose photodissociation bands lie at shorter UV wavelengths (such as CO, see Figure~\ref{fig:timescales}, top right). This leads to a considerable change in the unshielded CO photodissociation timescale for the $\beta$ Pictoris gas disk; Figure~\ref{fig:timescales} (bottom right) shows that photodissociation at the clump radial location ($\sim$85 AU) has a significant contribution by the strong UV field of $\beta$ Pictoris as well as by the ISRF \citep[where the latter would instead dominate for a typical A5V stellar model, ][]{Kamp2000}. 

We therefore revise the unshielded CO photodissociation lifetime from 120 yrs \citep{Kamp2000, Dent2014} down to 42 yrs. 
We note that dust shielding is negligible for optically thin dust in debris disks \citep[e.g.][]{Kamp2000}, and H$_2$, if at all present, should have a low enough column density not to provide shielding either \citep{Matra2017a}. CO is however abundant enough that self-shielding cannot be neglected \citep{Matra2017a}. When this is taken into account, the CO lifetime increases from 42 to $\sim$105 yrs. 
This is accounted for in all calculations involving CO. 
We neglect self-shielding and shielding by CO for all other molecules due to their unknown spatial distributions and shielding functions. Neglecting self-shielding should be a good approximation given the much lower predicted abundance of molecules other than CO.

Overall, Figure~\ref{fig:timescales} (left) shows that the harsh UV radiation field around $\beta$ Pic causes photodestruction of the vast majority of volatile species to take place in timescales of days to months. 
Therefore, photodissociation alone can explain most of our upper limits, as we confirm more quantitatively in \S\ref{sect:pred}, where we make predictions for future surveys (see Figure~\ref{fig:molpreds}, right).
N$_2$, CO and CN are by far the longest lasting molecules with unshielded photodissociation timescales of 96, 42 and 16 years, at least an order of magnitude longer than for any other molecular species.
This means that the only observable molecule with a survival timescale similar to that of the detected CO is CN, which we therefore focus on for the rest of this work. In Solar System comets CN gas is produced mainly by photodissociation of HCN gas \citep{Fray2005}; we will here assume that the same applies for exocomets around $\beta$ Pictoris. 

One potential caveat to this assumption is that although for simplicity we have neglected gas-phase chemical production/destruction pathways other than photodissociation, CN could be destroyed by reaction with atomic oxygen \ion{O}{1}, which has been detected in the disk around $\beta$ Pictoris \citep[e.g.][]{Brandeker2016, Kral2016}. Unfortunately, the detection being unresolved spectrally and spatially combined with excitation and optical depth effects make number density estimates at the locations where CN is expected to lie very uncertain. Nonetheless, combining a number density value of $2\times10^2$ cm$^{-3}$ at 85 AU from the best-fit dynamical evolution model of \citet{Kral2016} with a reaction rate of $5\times10^{11}$ cm$^{3}$ s$^{-1}$ from the KIDA database yields a reaction timescale of $\sim$3 years. This is shorter than the CN photodissociation timescale (16 years), indicating that reaction with oxygen could introduce an uncertainty of a factor of a few to our calculations, increasing the inferred HCN/CO outgassing ratio. Given the uncertainty on the oxygen abundance, however, we opt to neglect this destruction pathway of CN in this work.

\subsection{The effect of molecular excitation including fluorescence}
\label{sect:exc}
The only unknowns remaining to link the observed $F_{\rm CN\ N=2-1,\ J=5/2-3/2}$ / $F_{\rm CO\ J=2-1}$ flux ratio to the $\dot{N_{\rm HCN}^+}$/$\dot{N_{\rm CO}^+}$ ratio of the HCN and CO outgassing rates are the fractional populations of the upper levels $x_{\rm N=2,\ J=5/2,\ CN}$ and  $x_{\rm J=2,\ CO}$ of the observed line transitions. \citet{Matra2015} showed that, at least for CO, the assumption of local thermodynamic equilibrium (LTE) is unlikely to apply in the low gas-density environments present in debris disks. This prediction was confirmed by CO line ratio observations in $\beta$ Pictoris \citep{Matra2017a}, and observations of atomic gas species \citep{Kral2016,Kral2017}. 
To determine the CO and CN level populations we therefore carry out a full non-local thermodynamic equilibrium (NLTE) calculation which includes solving the equations of statistical equilibrium for both molecules. NLTE level populations depend on the local radiation field within the disk, and on the local density of the dominant collider species (here taken to be electrons) $n_{e^-}$ and the local kinetic temperature $T_k$. We direct the reader to \citet{Matra2015} for details.

\begin{figure}
\hspace{-6.5mm}
  \includegraphics*[scale=0.31]{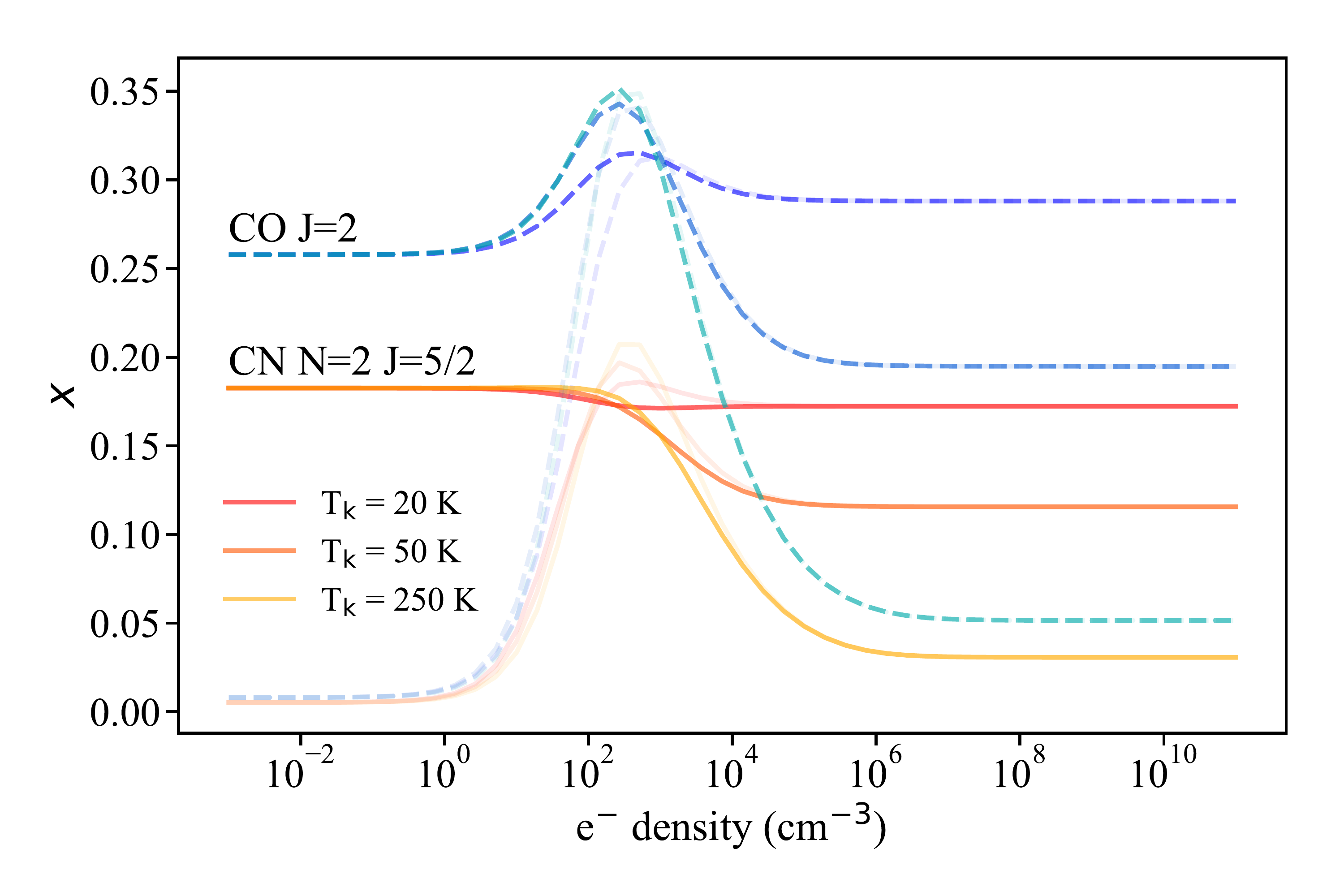}
\vspace{-8mm}
\caption{Fractional populations $x$ of the upper level of the observed transitions for CO (dashed lines) and CN (solid lines), assuming the radiation field felt by molecules at 85 AU from the star. Fainter lines show the case where fluorescence is not taken into account. In general, the populations depend on the density of the dominant collider species \citep[here electrons, x axis,][]{Matra2015}, and the kinetic temperature of the gas (colours). Increasingly darker colours indicate increasingly lower temperatures.}
\label{fig:CNvsCOfracpops}
\end{figure}

Thin dashed and solid lines in Figure~\ref{fig:CNvsCOfracpops} show the resulting dependence of the CO and CN upper level fractional populations $x$ on the electron density and temperature in the gas disk, when applying our previous NLTE model \citep{Matra2015}. Upper level populations vary between two limiting regimes, where excitation of the molecules is dominated by either radiation (far left, low $n_{\rm e^-}$) or collisions (i.e. LTE, far right, high $n_{\rm e^-}$). 
As expected, the calculated excitation of the considered CO and CN levels drops by several orders of magnitude in the radiation-dominated regime. This is because the model so far included only low-energy rotational transitions at far-IR/mm wavelengths, which can only be excited by the faint CMB and disk continuum emission.

In Appendix \ref{app:a}, we expand this NLTE excitation model to include vibrational and electronic as well as rotational levels for both CO and CN. This allows us to account for the effect of fluorescence, which in low-density regimes such as Solar System comets and exocometary gas disks, where radiation can dominate excitation, becomes an important excitation mechanism. Fluorescence works through absorption of stellar UV (IR) photons from low rotational levels in the ground electronic and vibrational state to excited electronic (vibrational) levels. This is followed by rapid relaxation to the ground electronic and vibrational state, but to higher rotational levels than the molecule started from. As we can see in Fig. \ref{fig:rotpops} (e.g. green versus red lines), 
this effectively leads to enhanced population of rotational levels above ground, where this can be orders of magnitude higher compared to when fluorescence is not considered. We note that fluorescence matters only in the low-density, radiation-dominated regime (left in Fig. \ref{fig:CNvsCOfracpops}, thick versus thin lines), whereas populations in the LTE, collision-dominated regime (right in Fig. \ref{fig:CNvsCOfracpops}) remain unchanged. The extent to which fluorescence impacts molecular excitation depends on the stellar flux received by the molecule and its spectral shape from UV to IR wavelengths.

In general, we find that taking fluorescence into account is crucial when using NLTE excitation models to derive molecular gas masses from observed fluxes. This impacts previous CO mass measurements where this effect was not taken into account; while the exact impact depends on the UV and IR flux received by a molecule which is different for each observed system, the overall effect of the enhanced excitation will be to lower the upper limits in the range of NLTE-derived masses. The total CO mass around $\beta$ Pic remains largely unchanged as it is tightly constrained observationally through the J=3-2/J=2-1 CO line ratio. For completeness, we report an updated value of $3.6^{+0.9}_{-0.6}\times10^{-5}$ M$_{\oplus}$, calculated in the same way as \citet{Matra2017a}. Including fluorescence will also lower the electron densities derived in Fig. 11 of that work by a factor of a few, bringing them closer to the model predictions of \citet{Kral2016}, and flatten their radial distribution in the inner regions of the CO disk. 

Taking fluorescence at the clump location into account, we find that the ratio in fractional upper level populations between CN and CO varies between 0.49 and 0.71, with the lowest value for electron densities of 10$^2$-10$^3$ cm$^{-3}$ and temperatures above 50 K. We note that this electron density - temperature dependence of the ratio between CN and CO fractional upper level populations is much weaker than for each individual upper level population, due to the similarity in the structure of rotational energy levels between CO and CN.

\subsection{Deriving the HCN/(CO+CO$_2$) ratio of outgassing rates}
\label{sect:hcnovco}



Plugging CN/CO ratios of upper level populations, photodissociation timescales and observed fluxes into Eq. \ref{eq:releaseratetoflux} allows us to derive an upper limit on the HCN/CO ratio of outgassing rates, which will carry the same dependence on electron density and temperature as the ratio between the CN and CO fractional populations mentioned above (coloured lines in Fig. \ref{fig:CNvsCOfracpops}). The lowest CN/CO ratio in upper level fractional populations sets the most conservative upper limit of 2.5\% on the HCN/CO ratio of outgassing rates.

An important point to make is that CO gas is also produced via rapid photodissociation of CO$_2$ gas. This implies that if CO$_2$ is being released from cometary ice together with CO, which we deem likely (see discussion in \S\ref{sect:depl}), the CO gas observed has contributions from both the CO and CO$_2$ ice reservoirs. From now on, we therefore consider our upper limit of 2.5\% to trace the HCN/(CO+CO$_2$) rather than the HCN/CO ratio of outgassing rates.
With this limit, in \S\ref{sect:disc} we draw the comparison between $\beta$ Pictoris and Solar System comets, and discuss the origin of a potential abundance discrepancy. 

\section{Discussion}
\label{sect:disc}


\subsection{HCN/(CO+CO$_2$) outgassing rates in $\beta$ Pictoris exocomets vs. Solar System comets}
\label{sect:depl}

\begin{figure}
\hspace{-6.5mm}
  \includegraphics*[scale=0.37]{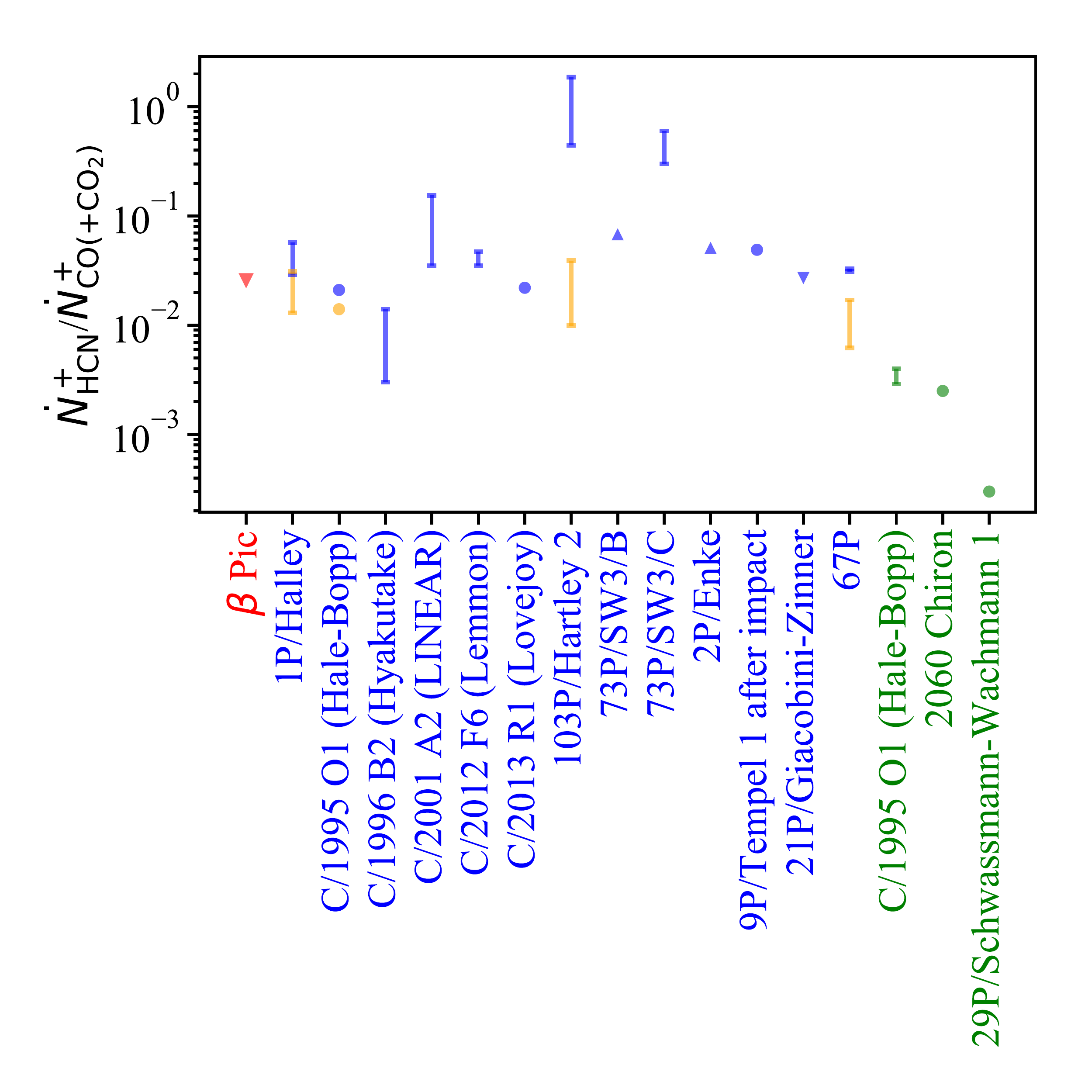}
\vspace{-8mm}
\caption{Comparison between our measured upper limit on the HCN/(CO+CO$_2$) ratio of outgassing rates in $\beta$ Pictoris (red) with Solar System comets. 
For Solar System comets, symbols represent the HCN/(CO+CO$_2$) outgassing ratios (orange, where available) or HCN/CO outgassing ratios (blue) reported by \citet{LeRoy2015} and references therein. We use filled circles where a single value is reported, and vertical bars where a range of values is reported. Upward and downward pointing triangles represent lower and upper limits, respectively. Green points are HCN/CO measurements for comets observed at large distances from the Sun \citep[Hale-Bopp at 4.6-6.8 AU, Chiron at 8.5-11 AU, and 29P at 5.8 AU,][]{Womack2017}, where different thermal outgassing rates are expected for CO and HCN.}
\label{fig:HCNoverCOoutgasrat}
\end{figure}

Figure~\ref{fig:HCNoverCOoutgasrat} shows our conservative upper limit of 2.5\% on the ratio of the HCN/(CO+CO$_2$) outgassing rate in $\beta$ Pic (red) compared to HCN/(CO+CO$_2$) and HCN/CO values measured in Solar System comets from \citet{LeRoy2015} and references therein (orange and blue symbols, respectively). We find that the upper limit we set in $\beta$ Pic is at the low end of the values derived for comets as they approach the Earth in our own Solar System.
Although our current upper limit per se does not yet indicate a significant depletion compared to Solar System values, we here consider how a low value (and, if confirmed in future, a true depletion) could be linked to either the outgassing mechanism itself, or to the intrinsic HCN/(CO+CO$_2$) abundance in the ice.  

The spatial distribution of CO provides an important clue about the gas release mechanism, as CO emission is peaked at a clump located $\sim$85 AU from the star \citep{Dent2014,Matra2017a} that coincides with a dust enhancement most clearly observed in mid-IR imaging \citep{Telesco2005}. This dust enhancement should be associated with higher collisional rates compared to the rest of the belt, and it strongly suggests both the CO and dust enhancements originate from collisions, directly or indirectly. 

The measured outgassing rates of molecular species (integrated over all grain sizes) should reflect the ice composition \textit{as long as both ice species are released as gas before they reach the bottom size of the collisional cascade} \citep{Matra2017b}. 
Whether this is the case depends on the details of the release process. Gas may be released during collisions through at least three mechanisms:
(1) direct collisional release of species that should have already sublimated at the temperature characteristic of the belt, but were trapped beneath a layer of refractories or within other ice matrices prior to the collision; 
(2) thermal sublimation following collisional heating due to e.g. hypervelocity impacts \citep[e.g.][]{Czechowski2007}, or (3) UV photodesorption of resurfaced ice, which was previously buried below a layer of refractories \citep{Grigorieva2007}. 

While CO is considered a supervolatile and is most likely fully lost by the time grains reach the blow-out size \citep{Matra2017b}, HCN is not and will not sublimate even at the temperature of the smallest grains in the $\beta$ Pic belt (which dominate the emission and lead to a directly measurable value of $\sim$86 K from the spectral energy distribution). Such temperature dependence on the relative outgassing rates is observed in Solar System comets, where sublimation drives the outgassing and causes the HCN/CO ratio of outgassing rates to decrease with increasing distance from the Sun \citep{Womack2017}. For comparison, we note that $\beta$ Pictoris is much more luminous than our own Sun \citep[$L_{\star}$=8.7 L$_{\odot}$,][]{KennedyWyatt2014}, and that we are observing exocometary CO at distances of $\sim$50-220 of AU, where blackbody temperatures are in the range $\sim$30-70 K. The same temperatures are attained in the Solar System at a distances of 19-39 AU from the Sun, i.e. from the orbit of Uranus out to the Kuiper belt. This means that $\beta$ Pic's exocomets are much colder than observed active comets in our Solar System, even compared to the distantly active ones mentioned above (observed at $\sim$5-11 AU from the Sun). This implies that we would expect CO but not HCN to sublimate from $\beta$ Pic's exocomets. The question is then whether thermal sublimation is the dominant gas release mechanism in the $\beta$ Pic disk.


Given the strong UV field from the central A-type star, UV photodesorption of ice resurfaced after collisions is a promising mechanism to release ice species less volatile than CO to the gas phase. The calculations of \citet[][]{Grigorieva2007} show that the timescale for complete UV photodesorption of a pure H$_2$O ice grain is shorter than the collision timescale for grains below $\sim$20 $\mu$m in size. Assuming HCN has a similar photodesorption yield as water, and neglecting any trapping within a refractory layer, this implies that all of the HCN ice will be released from grains before reaching the smallest size in the collisional cascade. A similar argument applies to CO$_2$, which has a volatility in between that of CO and HCN \citep{Matra2017b}. UV photodesorption is therefore most likely at play, and if dominant that would imply that HCN and CO$_2$ should be released in the gas phase together with CO, at rates reflecting their ice abundances.

Even if HCN does not photodesorb and survives in the ice-phase on the smallest grains in the cascade, it can be thermally sublimated as these unbound grains collide on their way out of the system \citep{Czechowski2007}. This is because stellar radiation pressure accelerates them to sufficiently high velocities for collisions to provide sufficient heating for sublimation. Application of this model to the $\beta$ Pic disk shows that this process can release ice (of unspecified composition) in the gas phase \citep{Czechowski2007}, with gas production rates of the same order as those estimated for CO in the belt. Then, similarly to UV photodesorption, and as long as all ice species are fully sublimated in each impact, high velocity collisions of unbound grains would also produce gas release rates reflecting ice abundances.

In summary, a low HCN/(CO+CO$_2$) ratio of outgassing rates could be attributed to the gas release mechanism only if this is driven by low velocity collisions, as this induces release of trapped supervolatiles alone. However, we would expect outgassing of all molecules independent of their sublimation temperatures if UV photodesorption and/or high velocity collisions with unbound grains are the dominant outgassing mechanisms. 
Detailed modeling results \citep{Grigorieva2007,Czechowski2007} show that these processes are most likely at play for small grains in the $\beta$ Pic belt. Then, unless most CO is being released through low velocity \textit{non-catastrophic} collisions between large bodies at the top of the cascade \citep[which are not taken into account by our model, see][]{Matra2017b}, we would expect ratios of outgassing rates to reflect cometary ice abundances. These can then be directly compared with relative abundances from Solar System comets, measured as they come close enough to the Sun for their ratios of outgassing rates to reflect ice compositions.


In the scenario where outgassing in the $\beta$ Pic belt is dominated by UV photodesorption or high velocity collisions of small grains, our upper limit on the HCN/(CO+CO$_2$) ratio of outgassing rates for $\beta$ Pic's exocomets suggests an HCN/(CO+CO$_2$) ice abundance ratio that is at the low end of the range observed in Solar System comets, although still not exceptionally low. 
Future, deeper observations (Sect. \S\ref{sect:pred}) are required to conclusively determine whether the abundance of $\beta$ Pic's exocomets is truly abnormal compared to Solar System comets.

Determining the abundance of HCN in exocomets is especially interesting because of its connections to the origins of life chemistry \citep[e.g.][]{Powner2009}. The large scatter in HCN abundances observed amongst Solar System comets \citep[][and references therein]{Mumma2011}, together with observation of cyanides in protoplanetary disks \citep{Oberg2015, Guzman2017, Huang2017}, indicate that cometary cyanides may not be directly inherited from the ISM, but undergo active chemical processing during the epoch of planet formation. Whether the endpoint of this chemistry is different for different planetary systems is yet unclear and can only be resolved by future, deeper observations of exocometary cyanides.

\subsection{Predictions for molecular surveys of gas-bearing debris disks}
\label{sect:pred}

\begin{figure*}
 \hspace{-3mm}
  \includegraphics*[scale=0.4]{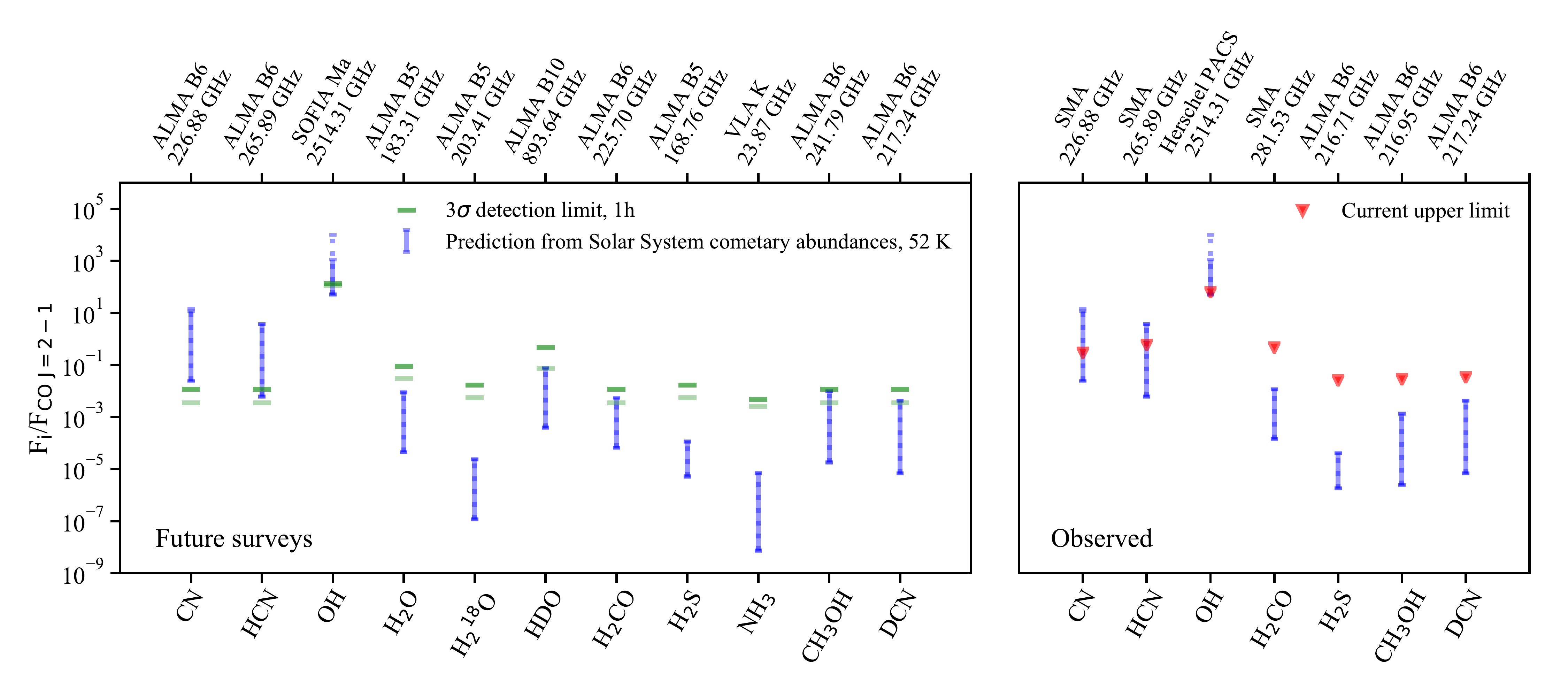}
\vspace{-8mm}
\caption{\textit{Left:} Predicted flux ratios compared to the detected CO J=2-1 line in the $\beta$ Pic disk, for Solar System cometary species observable at far-IR to radio wavelengths. For each species, vertical bars represent the expected flux ratios for the range of abundances with respect to CO observed in Solar System comets. Dashed vertical bars assume optically thin emission, whereas solid vertical bars take into account the optical thickness of the line (where this is only significant for OH). LTE excitation is assumed for all species, with a temperature of 52 K. The transition with the best `effective S/N' (see main text) is selected for each species. Green horizontal bars show the 3$\sigma$ threshold achievable with available facilities (SOFIA, ALMA, and the VLA) for 1h of integration and typical weather. Lighter green bars assume unresolved emission (best case scenario), whereas darker ones consider the loss in sensitivity expected for resolving the disk if it has the same on-sky extent as the CO disk, for the lowest spatial resolution of the instrument at the frequency in question (worst case scenario). Labels on the top x axis show the instrument and frequency best suited for detection of each species. \textit{Right:} The same predictions are now compared to the upper limits (red triangles) achieved in this work for molecular transitions observed with the SMA, ALMA and \textit{Herschel}.
}
\label{fig:molpreds}
\end{figure*}


To guide future observations, we
use Eq. \ref{eq:releaseratetoflux} to calculate expected line flux ratios around $\beta$ Pictoris for all common cometary species taking into account the range of abundances observed in the Solar System. We make two main assumptions. First, we assume all species are co-located with CO, leading to similar spatial distributions and excitation conditions. Since the bulk of the CO emission originates from the SW clump at 85 AU from the star, we use the stellar radiation field at 85 AU as representative of the bulk of the gas disk. Second, we calculate the excitation of all species using the LTE approximation \citep[apart from CO itself, where the level populations are very well constrained by the existing J=3-2/J=2-1 line ratio,][]{Matra2017a}. Figure~\ref{fig:CNvsCOfracpops} shows that the LTE assumption does not affect the calculated populations of CN and CO by more than a factor of a few for low-energy levels, but we cannot exclude that this may have a significant effect for other molecules and higher transitions. We vary the gas kinetic temperature between 52 K \citep[as predicted by thermodynamical models at the clump location,][]{Kral2016} and 210 K \citep[as the upper limit estimated through the resolved CO scale height at the clump location,][]{Matra2017a}.

We focus on molecular transitions observable with currently available facilities at far-IR to radio wavelengths, namely the Stratospheric Observatory for Infrared Astronomy (SOFIA), ALMA and the Karl G. Jansky Very Large Array (VLA). We select the best observable transition for each temperature and molecule to be that with the highest `effective S/N'. We construct this by dividing a transition's flux ratio with respect to CO (as calculated above) by the sensitivity of each telescope facility considered, at the frequency of the given transition. In addition, we consider the telescope's resolution by dividing this effective S/N by $\sqrt{N_{\rm beams}}$, where $N_{\rm beams}$ is the number of resolution elements covering the gas disk. As this acts to penalize facilities that over-resolve the gas disk, we choose a resolution closest to the CO disk's on-sky diameter \citep[$\sim16.5\arcsec$][]{Matra2017a} for the ALMA and VLA interferometers. This is a conservative approach as the spectral stacking technique described in \S\ref{sect:filtering} \citep[][Loomis et al. in press]{Matra2015} should significantly alleviate the sensitivity loss caused by over-resolving the disk.

Having selected the best transition for each molecule and temperature, we show its range of fluxes expected for Solar System cometary abundances in Figure~\ref{fig:molpreds}, left (vertical bars). We only show predictions for a single temperature of 52 K, as we find that the maximum effective S/N achievable for most species is largely independent of temperature (since e.g. for higher temperatures, higher transitions will emit more strongly and can then be selected for observations).
These predictions can then be compared with each telescope's 3$\sigma$ detection thresholds (green bars), where these also take into account that the disk will be resolved (darker as opposed to lighter bars for unresolved emission). We find CN (as expected), but also HCN and OH to be promising molecules for detection around $\beta$ Pic. This is largely due to their transitions having Einstein A coefficients orders of magnitude larger than CO, compensating for their lower gas masses caused by their shorter photodissociation timescales. 

Since the predicted line fluxes can, in a few cases, be up to orders of magnitude higher than CO, we need to consider that these transitions may be optically thick, particularly since CO itself has modest optical depth at the clump location ($\tau_{\rm CO\ J=2-1}\sim$0.12). To estimate optical depth, we first take the CO column density at the clump location as determined through J=3-2/J=2-1 line ratio observations \citep{Matra2017a}. Then, we scale this by the abundance ratio of molecule $i$ with respect to CO in the gas, to find the column density of molecule $i$.
This can be combined with LTE fractional populations to solve for the optical depth through its definition \citep[Eq. 3,][]{Matra2017a}, scaling the velocity width of the line to the Doppler-broadened width of the molecule in question. We find that the predictions, when accounting for optical depth effects (solid as opposed to dotted vertical bars in 
Figure~\ref{fig:molpreds}), change significantly only for the OH $^2\Pi_{3/2}\ \rm{J}=5/2^+-3/2^-$ line at 2.51 THz.

Overall, if cometary volatiles are being released at rates reflecting their ice composition, and if their composition reflects that of Solar System comets, we expect CN and HCN to be readily detectable with ALMA around $\beta$ Pictoris. Given that CN photodissociation takes place through photons of a similar wavelength to those causing CO photodissociation, we expect the CN/CO flux prediction to hold for exocometary gas disks around different stars as well.

An interesting result of our calculation is that OH detection should be possible with SOFIA. This encouraged us to check the \textit{Herschel} Science Archive, where we found unpublished PACS observations of $\beta$ Pictoris covering the OH line. We extracted the point-source-corrected spectrum from the central 7$\arcsec$ spaxel of the PACS Level 3 data product\footnote{\textit{Herschel} Observation ID: 1342198170}, and scaled it to the total flux of the central 3x3 spaxels. Then, we carried out continuum subtraction through a second order polynomial, leading to a measured RMS noise level of 0.24 Jy. Given the coarse spectral resolution of the data (295.2 km/s at 119.3 $\mu$m), we assume the line is unresolved to set a 3$\sigma$ upper limit on the integrated line flux of 2.1$\times$10$^2$ Jy km/s. We find that this is already at the very low end of our predicted range. However, given the significant optical thickness of the line, our prediction is very sensitive to the choice of OH temperature or spatial distribution, which may be oversimplified here.

We find detection of rotational lines from the H$_2$$^{18}$O water isotopologue, hydrogen sulfide (H$_2$S) and ammonia (NH$_3$) to be very unlikely with current instrument capabilities, although ammonia may be detectable in future with the sensitivity improvement brought by the Next Generation Very Large Array\footnote{\url{http://library.nrao.edu/public/memos/ngvla/NGVLA_17.pdf}}. On the other hand, we can already probe Solar System cometary levels for molecules such as the main isotopologue of water (H$_2$O), heavy water (HDO), formaldehyde (H$_2$CO) and methanol (CH$_3$OH), which could be detected if their respective ices are at the high abundance end of Solar System composition compared to CO. We note that for these species, like OH, the short photodissociation timescale would cause an even more pronounced clump compared to CO observations (an aspect which we have neglected here for simplicity). This would mean that molecular emission would have a higher optical depth than predicted here. At the same time, the more compact emission would be less over-resolved compared to CO, leading to higher spatially integrated S/N, and lower upper limits as derived here through our spectro-spatial filtering technique. 
Finally, we note that an increased optical depth \textit{to UV photons} could also prolong the survival lifetime of these molecules, leading to more favourable conditions for detection in future surveys.


\section{Conclusions}
\label{sect:concl}

We have presented the first molecular survey of the exocometary gas disk around $\beta$ Pictoris with the SMA, and combined it with archival ALMA data to obtain coverage of 10 different molecular species, namely CO, CN, HCN, HCO$^+$, N$_2$H$^+$, H$_2$CO, H$_2$S, CH$_3$OH, SiO and DCN. We reported continuum and CO J=2-1 detections and used upper limits on all other species as the basis to expand an exocometary release model beyond CO. We obtain the following conclusions:

\begin{enumerate}
\item The (optically thin) line fluxes for a given molecule depend on its exocometary outgassing rate, its photodissociation lifetime (neglecting other gas-phase chemical reactions), its intrinsic line strength, and the fractional population of the upper level of the transition (Eq. \ref{eq:releaseratetoflux}). Knowledge of this upper level population requires NLTE excitation modelling which we here expand to include the effect of fluorescence. We find that the latter produces orders of magnitude higher excitation in the low gas density, NLTE regime, and needs to be taken into account for robust derivation of gas masses from observed line fluxes.
\item CO is the longest-lived cometary molecule against photodissociation that is observable at millimeter wavelengths. This long lifetime explains, in part, why it is the only one that has been detected so far. Other observable molecules have survival lifetimes several orders of magnitude below that of CO, except for CN. This makes CN the most promising molecule for detection in debris disks after CO. 
\item CN is a daughter molecule, which we assumed to be produced mostly through photodissociation of HCN gas released from the ice phase. CO gas on the other hand, is both a parent and daughter species produced by CO$_2$ photodissociation. Therefore, we used our upper limit on the CN/CO flux ratio to derive an upper limit on the HCN/(CO+CO$_2$) ratio of outgassing rates of 2.5\%. This is still consistent with, although below most of the values measured in Solar System comets as observed near Earth. 
If deeper CN and/or HCN observations push this HCN/(CO+CO$_2$) outgassing rate upper limit down to a significant depletion compared to Solar System comets, we show that this may be caused by either the outgassing mechanism or a true depletion in the ice abundance. 
\item Outgassing ratios of molecular species reflect their ice compositions as long as ices cannot survive on grains down to the blow-out size in the collisional cascade and be removed from the system. This is likely the case in $\beta$ Pictoris, as 1) collisions will allow ice layers to resurface, where they become subject to rapid UV photodesorption \citep{Grigorieva2007}, and 2) the smallest unbound grains will undergo hypervelocity collisions on the way out of the system leading to vaporization of any ice that may have survived on these grains \citep{Czechowski2007}. In general, these processes will allow release of molecules less volatile than CO that would otherwise survive in the ice phase on the smallest grains and be dynamically removed by radiation pressure from the central star.

\item Then, unless non-catastrophic collisions at the top of the cascade are the dominant gas release mechanism, a low HCN/(CO+CO$_2$) outgassing rate is attributed to a low HCN/(CO+CO$_2$) abundance in the exocometary ice. If confirmed by deeper observations, this would indicate that active cyanide chemistry during planet formation produces a wide variety of cometary compositions across planetary systems.

\item We present predictions for future detection of cometary species around $\beta$ Pictoris, assuming Solar System abundances and LTE excitation. We find that ALMA should readily detect CN and HCN line emission, even for the most HCN-poor Solar System abundances. Other species such as H$_2$O, HDO, H$_2$CO, CH$_3$OH and DCN may be detectable, depending on their specific composition, spatial distribution and excitation conditions.
\item Finally, we report \textit{Herschel} archival upper limits on the OH $^2\Pi_{3/2}\ \rm{J}=5/2^+-3/2^-$ line at 2.51 THz that are already at the very low end of the range predicted assuming LTE and Solar System cometary H$_2$O abundances. The line is optically thick and hence the flux prediction is sensitive to the assumed on-sky distribution and temperature of OH gas, which are likely oversimplified here. Further modelling taking these effects into account is needed to evaluate whether this upper limit indicates an underabundance of water in $\beta$ Pic's exocomets, and whether SOFIA may be able to detect OH from photodestruction of water in exocometary gas disks.

\end{enumerate}



\acknowledgments
The authors are grateful to A. Brandeker and W.-F. Thi for providing the spectrum of the star and the complete rate matrix of CO including vibrational and electronic levels. The authors would also like to thank Quentin Kral and Gianni Cataldi for helpful notes on the submitted manuscript. LM acknowledges support from the Smithsonian Institution as a Submillimeter Array (SMA) Fellow. KI\"O acknowledges support from the Simons Collaboration on Origins of Life (SCOL).
The Submillimeter Array is a joint project between the Smithsonian
Astrophysical Observatory and the Academia Sinica Institute of Astronomy and
Astrophysics and is funded by the Smithsonian Institution and the Academia
Sinica.
This paper makes use of ALMA data ADS/JAO.ALMA\#2012.1.00142.S. ALMA is a partnership of ESO (representing its member states), NSF (USA) and NINS (Japan), together with NRC (Canada), NSC and ASIAA (Taiwan), and KASI (Republic of Korea), in cooperation with the Republic of Chile. The Joint ALMA Observatory is operated by ESO, AUI/NRAO and NAOJ. \textit{Herschel} is an ESA space observatory with science instruments provided by European-led Principal Investigator consortia and with important participation from NASA.



\facility{SMA, ALMA, \textit{Herschel}}.
\software{
MIR, miriad \citep{Sault1995},
CASA \citep{McMullin2007},
Matplotlib \citep{Hunter2007} 
}

\bibliographystyle{apj}
\bibliography{lib}

\appendix

\section{Solving the statistical equilibrium for CO and CN including fluorescence}
\label{app:a}

We direct the reader to \S2 in \citet{Matra2015} for a detailed explanation of how to calculate the rotational excitation of a molecule (using the example of CO) for a given radiation field $J$, density of collisional partners  $n_{\rm coll}$ and kinetic temperature $T_{\rm k}$. In summary, the total rate of radiative and collisional transitions populating each of N$_{\rm lev}$ energy levels is equated to the total rate of transitions depopulating the level. The statistical equilibrium calculation involves solving this system of N$_{\rm lev}$ linear equations to obtain the fractional population $x$ of each energy level (defined as the number of molecules in that level divided by the total number of molecules). 

We here expand this calculation to include not only rotational transitions (lowest energy, corresponding to far-IR/mm wavelengths), but also vibrational (near/mid-IR) and electronic (UV/optical) transitions. This inclusion is important in the presence of strong UV/IR radiation fields such as in our case of optically thin gas around an early-type star. Although in this work we are only interested in the population of rotational levels, as it affects our observed mm lines, absorption of UV photons is important as well. The latter can excite the molecule to high electronic states followed by rapid relaxation to the ground electronic state. However, this relaxation takes place via cascading through excited vibrational and/or rotational states (i.e. fluorescence), and this may significantly affect the rotational populations we are interested in \citep[see Figure 1 in][for a schematic view of the process for the CO molecule]{Thi2013}.

For CO, we obtain radiative transition rates from \citet{Thi2013} and references therein. For CN, pure rotational radiative rates were obtained from \citet{Klisch1995} using the dipole moment of \citet{ThomsonDalby1968} through the LAMDA\footnote{\url{http://home.strw.leidenuniv.nl/~moldata/}} database \citep[][]{Schoier2005}, whereas all other rovibrational/electronic radiative rates were obtained from \citet{Brooke2014} through the ExoMol\footnote{\url{http://www.exomol.com/}}  database \citep{TennysonYurchenko2012}. This resulted in our compilation of all allowed radiative transitions between any of the 2 (3) lowest electronic levels, 9 (6) lowest vibrational levels, and 30 (20) lowest rotational levels for CO (CN). For CN, each rotational level (N) apart from N=0 is divided into 2 closely spaced fine-structure (J) levels for all electronic and vibrational levels; in addition, J levels in the A$^2\Pi$ state  (first electronic state above ground) are further subdivided into 2 levels due to $\Lambda$-doubling. We neglect hyperfine splitting of the lines, which acts to create sub-transitions that are mostly blended within the Keplerian velocity structure of each fine-structure (J) transition observed.

The gas kinetic temperatures in gas-bearing debris disks are expected to be of order a few tens of K, as predicted by thermodynamical models \citep{Kral2016} and as suggested by the CO excitation temperatures observed so far \citep[e.g.][]{Matra2017a, Hughes2017}.
This means that collisional population of excited vibrational and electronic energy levels (corresponding to temperatures of thousands to tens of thousands of K, respectively) can be neglected. Therefore, we only include collisional transitions between the rotational levels of the ground vibrational and electronic state, using rates from \citet{Dickinson1975} for CO, and from \citet{AllisonDalgarno1971} through the LAMDA database for CN.

Calculating the absorption and stimulated emission rates between any two levels of the molecule requires knowledge of the radiation field $J$ at a frequency $\nu$ corresponding to the energy difference between the two levels. Given the inclusion of UV and near-IR as well as mm transitions, knowledge of the radiation field from UV to mm wavelengths as felt by a molecule at a given location in the gas disk is required. We take into account contributions from the cosmic microwave background (CMB),  the star (using the same spectrum as in \S\ref{sect:lifetimes}), and the disk's dust continuum emission. For the latter, we use the best-fit 3D model to ALMA millimeter continuum observations \citep[][]{Matrainprep} to measure the radiation field within the disk itself, as described in Appendix B of \citet{Kral2017}; we then scale the total disk emission to other wavelengths using unresolved SED flux measurements. However, we find inclusion of the dust continuum radiation field to have a negligible effect on the excitation calculation. This is because the dust continuum is sufficiently faint that absorption of stellar UV photons and subsequent spontaneous emission take place on a shorter timescale than absorption and stimulated emission from IR dust continuum photons. As well as that, the CMB dominates excitation over the dust continuum for the low-lying rotational transition of interest here \citep[as found for Fomalhaut in ][Fig. 3]{Matra2015}.

\begin{figure}
\hspace{-5mm}
 \centering
  \includegraphics*[scale=0.5]{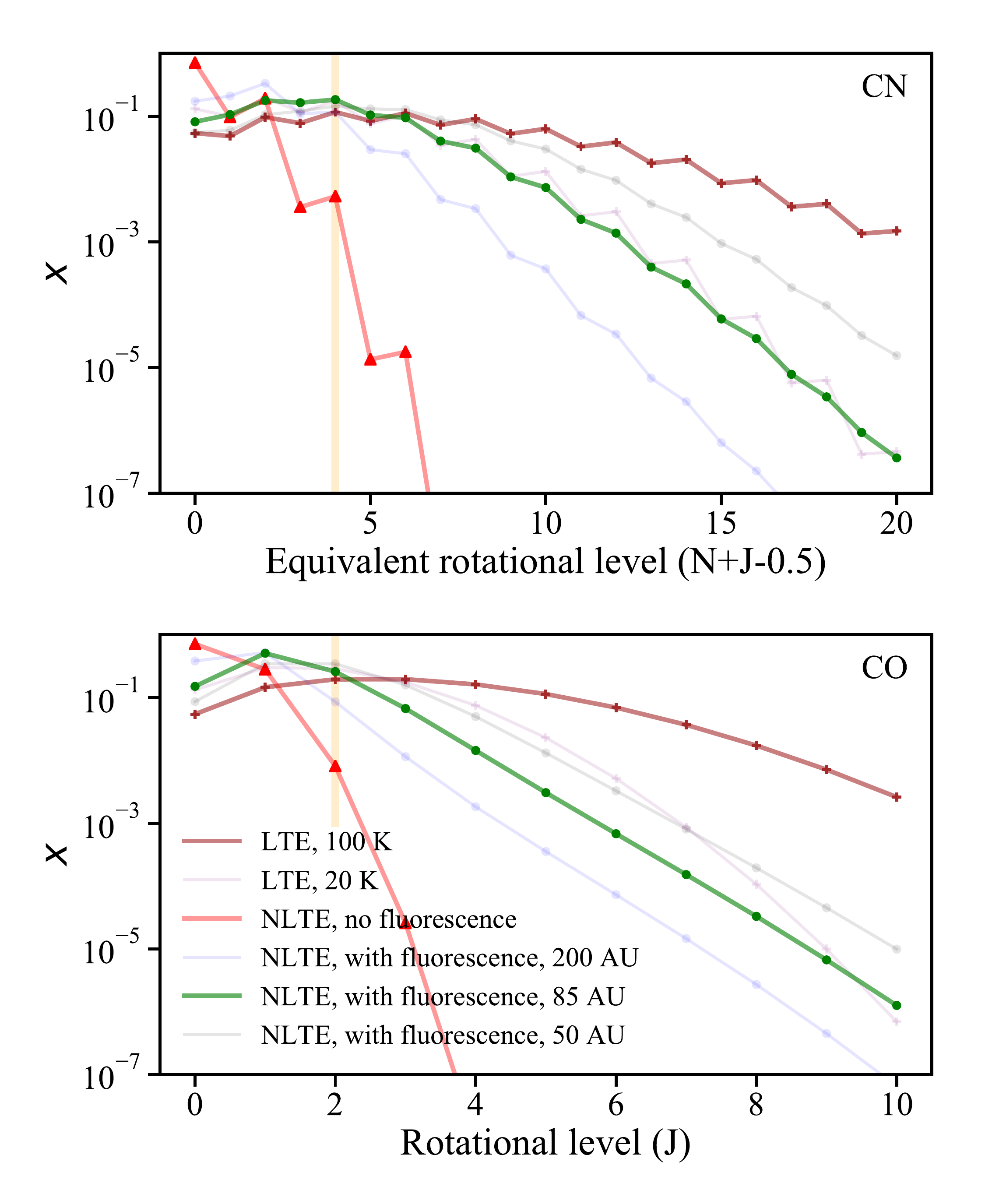}
\vspace{-6mm}
\caption{Fractional population $x$ of rotational energy levels of CO and CN for a variety of excitation conditions in the $\beta$ Pictoris gas disk. Brown and purple crosses are populations in the LTE regime, for kinetic temperatures of 20 and 100 K. Red triangles are populations in the NLTE, radiation-dominated regime, neglecting fluorescence from excited vibrational and electronic energy levels. Filled circles show how the result changes when including fluorescence, for the radiation field felt by a molecule at 50 (gray), 85 (green) and 200 (blue) AU from the star. The yellow band shows upper levels of the transitions observed in this work. Note that for CN, we label levels through a combination of the N and J quantum numbers, in a way that ensures CN and CO levels with similar energies are vertically aligned across the two panels.}
\label{fig:rotpops}
\end{figure}

Figure~\ref{fig:rotpops} shows the results of the statistical equilibrium calculation for the pure rotational ladder (ground vibrational and electronic state), which is relevant to millimeter observations such as described in this work. In particular, the fractional populations $x$ of levels N=2, J=5/2 for CN and J=2 for CO (highlighted by the orange line) are the ones displayed in Figure~\ref{fig:CNvsCOfracpops} as a function of the kinetic temperature and electron density of the gas. From that figure, it is clear that for each level, two limiting regimes are present \citep[in line with our expectation, see e.g.][]{Matra2015}; in the very low gas density regime (which we here denominate `far-NLTE') excitation by radiation dominates, whereas in the high gas density regime it is collisional excitation that dominates (LTE). In both limiting regimes the excitation becomes independent of the gas density; in the radiation-dominated regime, the excitation is also independent of temperature, whereas in LTE it is independent of the impinging radiation field. 

In the bottom (upper) panel of Figure~\ref{fig:rotpops}, we show how the rotational populations up to level J=10 (N=10) for CO (CN) behave in each limiting regime (LTE or far-NLTE). We vary the temperature (affecting the LTE limit) and the distance from the star (affecting the stellar flux and hence the fluorescence mechanism in the NLTE limit). We also compare the NLTE results where fluorescence is turned on or off. As expected, in LTE (+ symbols) higher levels are preferentially populated at higher temperature, and our statistical equilibrium results match the analytical calculation from the Boltzmann distribution. In NLTE, we find that turning on the stellar UV field and hence the fluorescence mechanism (filled circles) leads to a significant population of rotational levels above ground compared to the case where fluorescence is not considered (red triangles). As we would expect, high rotational levels are even more favoured compared to low levels as we move closer to the star and thus increase the effect of fluorescence.

\end{document}